\documentstyle[12pt,psfig,epsfig,graphicx,psfrag]{article}
\newcommand{\be}{\begin{equation}}
\newcommand{\ee}{\end{equation}}
\newcommand{\ba}{\begin{eqnarray}}
\newcommand{\ea}{\end{eqnarray}}
\begin{document}

\titlepage
\begin{flushright}
\end{flushright}
\vskip 1cm
\begin{center}
{ \Large \bf Continuous spin particles from a string theory}
\end{center}
\vskip 1cm

\begin{center}
J. Mourad \footnote{mourad@th.u-psud.fr}
\end{center}
\vskip 0.5cm
\begin{center}
{\it APC} \footnote{Unit\'e Mixte 
de Recherche du CNRS (UMR 7164).}, Universit\'e  Paris VII,\\
2 place Jussieu - 75251 Paris Cedex 05, France\\
and\\
{\it LPT} \footnote{Unit\'e Mixte 
de Recherche du CNRS (UMR 8627).}, B\^at. 210
, Universit\'e
Paris XI, \\ 91405 Orsay Cedex, France

\end{center}
\vskip 2cm
\begin{center}
{\large Abstract}
\end{center}

It has been shown that the massless irreducible 
representations of the Poincar\'e group
with continuous spin can be obtained from  a classical point
particle action which admits a generalization to a conformally
invariant string action.
The continuous spin string action
is quantized in the BRST formalism. 
We show that
 the vacuum carries a continuous spin representation 
of the Poincar\'e group and  that 
the spectrum is ghost-free.
\newpage

\section{Introduction}

In this article we study the string action
\be
S[X^\mu,h_{mn}]=\mu\int d^2\sigma \sqrt{-h}\sqrt{\Box X^\mu\Box X^\nu\eta_{\mu\nu}}.
\label{act}
\ee
It shares with the classical Polyakov action the two-dimensional
local reparametrization
and Weyl symmetries as well as the global target space Poincar\'e
invariance. 
It also introduces a mass scale $\mu$ similar to the
string tension.
It differs however in two important
aspects: it is non-polynomial and it is a higher derivative
action. 
The action (\ref{act}) was first considered by
Savvidy in \cite{Savvidy:2003dv,Savvidy:2003fx,Savvidy:2004bb} 
where it was conjectured to play
a role in the tensionless limit of the Nambu-Goto-Polyakov
action \cite{Schild:1976vq,Karlhede:1986wb,Sundborg:2000wp}, 
it can be also motivated
as a generalization of the point particle action 
\cite{Mourad:2004fg}
describing
a massless particle belonging to the continuous spin
representations of the Poincar\'e group 
\cite{Wigner:1939cj,wi,
chak,Abbott:1976bb,Zoller:1991hs,Brink:2002zx}.
The quantization of the action (\ref{act}), due to its
non-polynomiality and higher derivative nature,
is not straightforward. These two difficulties can be 
overcome if one introduces an auxiliary field, called $\Xi^\mu$
in the following and a Lagrange multiplier. 
The action becomes second order
and polynomial. More explicitly the classically equivalent action is
\be 
S[X^\mu,\Xi^\mu,h_{mn},\lambda]=
-\mu \int d^2\sigma \sqrt{-h}\left[h^{mn}\partial_m X^\mu
\partial_n \Xi^\nu \eta_{\mu\nu}
+\lambda(\Xi^2-1)\right].\label{ac2}
\ee
In turn, this action presents some difficulties:
its kinetic term has signature $(D,D)$ signalling potential ghosts
and the constraints that arise from this action are
bilocal constraints. It was shown in \cite{Mourad:2004fg} that one can
transform these bilocal constraints to local one at the expense of
breaking manifest Lorentz covariance.
The quantization can then be performed with the standard BRST
techniques \cite{Becchi:1975nq,Kato:1982im,Henneaux:1985kr}. 
It was initiated in \cite{Mourad:2004fg} where it was found
that the critical dimension of the theory is 28.
It this article, we pursue this quantization in more details
and find that:

\noindent
-The ground state carries a continuous spin representation of the
Poincar\'e group.  

\noindent
-The BRST-closed higher level states are all of zero norm. 

This implies that the only physical state 
is the ground state. 
In this respect it has some similarities with
the N=2 superstring \cite{GSW} with a finite number of 
physical states.
It differs however in that the signature of spacetime is
$(1,D-1)$.
Although the manifest Lorentz covariance was broken
to transform the constraints, the resulting spectrum is
relativistic. The absence of ghosts in the spectrum, in spite of
the $(D,D)$ signature of the sigma model (\ref{ac2}),
is due to the additional constraints contained
in (\ref{ac2}) and proves the consistency
of the free theory.

This  theory may prove to be useful
to explore some properties of string theory.
The states being all massless, it may be relevant to
some very high energy regime of string theory.
This remains to be explored. It is also of crucial importance
to study interactions
and to elucidate the issue of unitarity
in the resulting amplitudes. In this respect let us note
that consistent interactions of particles belonging to the
massless continuous spin representations are not known.
An interesting  possibility would be 
that  string theory is needed in  order to
get consistent interactions. The theory
may also be considered as a simple toy model
for the quantization of systems with higher 
derivatives or bilocal constraints.

The plan of this paper is the following. In Section 2 we 
briefly recall the derivation of the constraints. Section 3 is
devoted to the BRST quantization where  
we calculate the critical dimension and the normal ordering
constant.
The latter is crucial for the ground state to belong to the
continuous spin representation. In Section 4
we show that the higher level BRST-closed states are all of zero
norm. The proof that the spectrum is ghost-free 
relies heavily on the requirement that
the zero modes of the antighosts annihilate the physical states.  
Finally, Section 5 contains the conclusions.

\section{String action and associated constraints}

The classical point particle action \cite{Mourad:2004fg}
(for earlier treatments see \cite{Zoller:1991hs}, see also
\cite{Edgren:2005gq})
for the  dynamical variables
$x^\mu$, the space-time position of the particle, and $\xi^\mu$,
the ``internal " degree of freedom, 
\be
S[x^\mu,\xi,e,\lambda]=\int d\tau {\dot x.\dot \xi \over e}
+e\tilde \mu +\lambda(\xi^2-1),\label{ac}
\ee
gives rise to the constraints 
\be
p.q-\tilde \mu=0,\ \ \xi^2-1=0,\ \ \xi.p=0,\ \ 
p^2=0.\label{conts}
\ee
Here $p$ and $q$ are the conjugate momenta to $x$ and $\xi$
respectively. These constraints are precisely those used by Wigner
\cite{wi} to describe in  a manifestly covariant
way the massless continuous spin representation
\footnote{Recall that these representations have $p^2=0$ and
$M_{\mu\nu}p^\nu M^{\mu\alpha}p_\alpha=\tilde \mu^2$,
where $p^\mu$ and $M_{\mu\nu}$ are the generators of translations
and Lorentz transformations and $\tilde \mu$ is the nonvanishing
continuous spin parameter. Notice that $\tilde \mu$ is not
necessarily positive.}.
 The generalization of (\ref{ac}) to 
 a two-dimensional world-sheet is given by
\be
S[X^\mu,\Xi^\mu,h_{mn},\lambda]=
- \mu \int d^2\sigma \sqrt{-h}\left[h^{mn}\partial_m X^\mu
\partial_n \Xi^\nu \eta_{\mu\nu} +\lambda(\Xi^2-1)\right],
\label{sac}
\ee
where $h_{mn}$ is the two-dimensional metric and 
the two target space coordinates $X$ and
$\Xi$ depend on the two world-sheet coordinates $\sigma^0$ and
$\sigma^1$ with $\sigma^1=\sigma^1+2\pi$. The action is classically invariant under
reparametrizations and
Weyl rescaling of the metric. 
The equations of motion are
\ba
\Box \Xi^\mu=0,\ \Box X^\mu={2\lambda }\Xi^\mu,\label{eqm}
\ea
where $\Box={1 \over
{\sqrt{-h}}}\partial_m\sqrt{-h}h^{mn}\partial_n$.
The primary constraints obtained by varying with respect to $\lambda$
and $h_{mn}$ are
\ba
(\Xi^2-1)=0,\label{l1}\\ 
\partial_m X^\mu\partial_n\Xi_\mu+\partial_n X^\mu\partial_m \Xi_\mu
-h_{mn}\partial_l X^\mu\partial^l\Xi_\mu=0.
\ea

Using (\ref{eqm}) and (\ref{l1}) we get for $\lambda$
\be
(2\lambda)^2=(\Box X)^2,\label{l2}
\ee

The second equation in (\ref{eqm}) together with
(\ref{l2}) can be used to determine $\Xi$ in terms of $X$ as
\be
\Xi={\Box X \over \sqrt{(\Box X)^2}}.
\ee 
We can eliminate $\Xi$ from the action to get the
higher derivative action
\ba
S&=& - \mu
\int d^2\sigma \sqrt{-h}h^{mn}\partial_m X^\mu
\partial_n {\Box X_\mu \over \sqrt{(\Box X)^2}}\nonumber\\
&=& \mu \int d^2\sigma \sqrt{-h}
\sqrt{\Box X^\mu \Box X_\mu},
\ea
which is the form proposed in \cite{Savvidy:2003dv}.

It was shown in \cite{Mourad:2004fg} that, in the conformal gauge,
in addition to the primary constraints
\ba
{\cal H}_0&=&P.Q+\mu^2 \partial_1X.\partial_1\Xi=0\nonumber\\
{\cal H}_1&=&P.\partial_1X+
Q.\partial_1\Xi=0,\nonumber\\
\phi_1&=&\Xi^2-1=0\label{pri},
\ea
the string action gives rise to the secondary constraint
\be
P.\Xi(\sigma),
\ee
and the bilocal constraints
\be
(P(\sigma)-\mu \partial_1\Xi(\sigma)).(P(\sigma')+
\mu \partial_1\Xi(\sigma'))=0.\label{bil}
\ee
Here $P$ and $Q$ are respectively  the conjugate 
momenta to $X$ and $\Xi$. 
In the following we shall use the definitions
$P_R(\sigma)=P+\mu\partial_1\Xi,\ P_L(\sigma)=P-\mu\partial_1\Xi$
of the right and left momenta and similarly for $Q_{R}$
and $Q_{L}$ so that
the first two constraints in (\ref{pri}) become 
$P_R.Q_R=P_L.Q_L=0$. 

We start by analyzing the bilocal constraint (\ref{bil}) 
rewritten
as
$P_R(\sigma).P_L(\sigma')=0$.
Let $V_R$ be the vector space spanned by $P_R(\sigma)$ when
$\sigma$ varies from $0$ to $2\pi$ and
similarly for $V_L$, then $V_R$ and $V_L$ are orthogonal.
Let $p$ be the common zero mode of $P_R$ and $P_L$.
By taking the integral on both $\sigma$ and $\sigma'$ of the
constraint (\ref{bil}) we
get that $p^2=0$. All the string modes are thus massless.
Furthermore $p$ is contained in both $V_R$ and
$V_L$. 
Suppose that the only nonvanishing component of $p$ is $p^+$
\footnote{We are using here the light-cone coordinates
$V^{\pm}=(V^0\pm V^{D-1})/\sqrt{2}$.},
and split the spacelike and transverse indices  $i=1,\dots D-2$
into $a$ which belong to $V_R$ and $a'$ which belong to $V_L$.
We thus have
\be
P_L^{a}(\sigma)=0,\ a=1,\dots N,\quad
P_R^{a'}(\sigma)=0,\ a'=N+1,\dots D-2,\label{contl}
\ee
and since $p^+$ is in both $V_R$ and $V_L$ we also have
\be
P_L^{-}(\sigma)=0=P_R^{-}(\sigma).
\ee
The latter two constraints are equivalent to 
$\Xi^-= \xi^-$
and $P^-=0$, with $ \xi^-$ a constant zero mode.
We have thus transformed the bilocal constraints into local 
ones at 
the expense of breaking the manifest Lorentz covariance.

The remaining constraints are
\be 
P.\Xi=0,\quad \Xi^2-1=0.
\ee
We wish to transform them, using (\ref{bil}), into constraints 
involving only left
movers or right movers. We shall be able to do so except for a
zero mode which involves the sum of right moving and left moving
variables. 
We have
\be 
P={P_R+P_L \over 2},\quad 
\Xi'={P_R-P_L \over 2\mu}.\label{inter}
\ee
Let $\bar P_R$ and $\bar P_L$ be the nonzero mode parts of
$P_R$ and $P_L$ and define $\hat P_R$ and $\hat P_L$ by
$\hat P_R'=\bar P_R$  the integration constant being
chosen so that $\hat P_R$ has no zero mode.
Integrate the second equation in (\ref{inter}) as
\be
\Xi(\sigma)={\hat P_R-\hat P_L \over 2\mu}+{\xi },
\ee
Here $\xi$ is a constant $D$-vector.
Notice that  we have $ \hat P_R.\hat P_L=0$
and $ P_R.\hat P_L=0$.

Using the derivative of $\phi_1$, $\Xi'.\Xi=0$, we get
\be
P_R.\Xi=0, \quad P_L.\Xi=0,
\ee
which yield
\be
G={\xi.P_R}+{1\over 2\mu}\hat P_R.
P_R(\sigma)=0,
\ee
and
\be
\tilde G={\xi.P_L}-{1\over 2\mu}\hat P_L.
P_L(\sigma)=0,
\ee
Notice that $G$ and $\tilde G$ have the same zero mode
\be
G_0={\xi.p \over 2\pi}=-{p^+\xi^- \over 2\pi}.
\ee
We have thus obtained one left moving   and one right
moving
constraints. It remains to take into account 
the zero mode constraint contained in $\phi_1$. 
This is accomplished by
\ba
g={\xi^2}+{1\over 4\mu^2 }\int {d\sigma \over 2\pi}  
(\hat P_R^2+\hat P_L^2)-1=0.
\ea
Notice that it is the sum of left moving variables and right
moving ones.

\section{BRST quantization}

Recall that for  a  system with first class
constraints $G_i$ which  
form a Lie algebra $[G_i,G_j]=C_{ij}{}^{k}G_k$
the first step in the BRST quantization is the 
enlargement of the
Hilbert space by the introduction
of the ghosts $c^i$ and 
antighosts $b_j$ which verify $\{b_i,c^j\}=\delta_i^j,
\{c^i,c^j\}=\{b_i,b_j\}=0$. The 
  BRST charge is  defined by $Q=c^iG_j-{1\over
2}C_{ij}{}^kc^ic^jb_k$, and verifies $Q^2=0$. 
The physical states are in the kernel of $Q$ and
two states are equivalent if they differ by an exact state,
i.e. of the form $Q|\phi>$ for some $|\phi>$.

In string theory, the classical constraints
form a closed Lie algebra.  When one turns the dynamical 
variables into operators, an anomaly appears 
in the commutators of the energy-momentum tensor.
This anomaly has two contributions: the first is dependent
on the normal ordering constant in the energy-momentum 
tensor and the second is the central charge.
More precisely, if one defines the normal ordered 
matter energy-momentum tensor
by
$T^{(m)}={\pi\over \mu}:P_R.Q_R:-{2D\over
24}+a^{m}$, with $a^m$ a constant,
then it obeys the commutation
relations\footnote{We are using the canonical commutation
relations
$[X^\mu(\sigma),P^{\nu}(\sigma')]=
i\eta^{\mu\nu}\delta(\sigma-\sigma')$ and so on.}
\ba
[T^{(m)}(\sigma),T^{(m)}(\sigma')]&=&
2\pi i(T^{(m)}(\sigma)+T^{(m)}(\sigma'))
\delta'(\sigma-\sigma')\nonumber\\
&-&4\pi i a^m\delta'(\sigma-\sigma')+
2\pi i{c^m \over 12}\delta'''(\sigma-\sigma'),
\label{commu}
\ea
where $c^m$ is the central charge of the matter sector given by
$2D$.

The other constraints, for the right moving part,
are given by
\ba
P^{a'}_R(\sigma)&=&0,\ \label{constra1}\\
P^{-}_R(\sigma)&=&0,\ \label{constra2}\\
 G={\xi.P_R}+{1\over 2\mu }\hat P_R.
P_R(\sigma)&=&0
.\ \label{constra3}
\ea
One has also to add the zero mode constraints $g=0$. 
The constraints (\ref{constra1}-\ref{constra3})
commute among each other and with $g$ and they have conformal weights 
equal to one,
that is if we denote generically one of the constraints (\ref{constra1}-
\ref{constra2}) by $K$ then we have
\be
\left[T^{(m)}(\sigma'),K(\sigma)\right]=-2\pi iK'(\sigma)
\delta(\sigma-\sigma')
-2\pi i K(\sigma)\delta'(\sigma-\sigma').
\ee
We also have
\be
[T(\sigma),g]=- {i\over \mu} G(\sigma),
\ee
and
\ba
\left[T^{(m)}(\sigma'),G(\sigma)\right]&=&-2\pi iG'(\sigma)
\delta(\sigma-\sigma')
-2\pi i G(\sigma)\delta'(\sigma-\sigma')\nonumber\\
&+&i\delta(\sigma-\sigma')
{p^+\over 2\mu}P_R^-(\sigma).\label{com3}
\ea
Equation (\ref{com3}) means that $G$ is weakly a field of
conformal weight $1$.
Let the ghosts fields 
associated to the constraints
(\ref{constra1}-\ref{constra3}) and $T^{(m)}$ be denoted 
respectively by
$c_{a'},c_{-},d$ and $c$ and the corresponding antighosts by
$b^{a'},b^{-},e$ and $b$. All the ghosts except $c$ 
 have conformal
weights $0$ and  $c$ has the conformal weight $-1$.
Let the ghost and antighost 
associated to $g$ be denoted by $\gamma$ and $\omega$.
These are not fields, they have only zero modes.
The naive BRST charge $Q$ which results
when ignoring the anomalous terms in 
the commutation relations of the energy-momentum tensor 
reads
\ba
Q&=&\int {d\sigma\over 2\pi} 
\Big[c(\sigma)T^{(m)} +c_{a'}P^{a'}_R+
c_{-}P^{-}_R+ d G\Big]+\gamma g \nonumber\\
 &+&\int {d\sigma\over 2\pi} 
 :c(\sigma)\Big[{1 \over 2}T^{(c)}
+T^{(c_{a'})}+T^{(c_{-})}+T^{(d)} -{i\over 2\pi\mu}\gamma e 
-{ip^+\over 4\pi\mu}d b^-+a\Big]:
,\label{brst}
\ea
$T^{(c)}$ is the energy-momentum tensor of the
ghost system $c$ and $b$ and so on for the other terms in 
(\ref{brst}). The energy-momentum tensor of 
the weight $h$ ghost \footnote{Here $h$ is the conformal weight of
the antighosts $b_h$ which is the same as that
of the corresponding constraint; we use the conventions 
$\{c_h(\sigma),b_h(\sigma')\}=2\pi\delta(\sigma-\sigma')$.}  system is
given by
\be
T_h=-i:[h\partial(c_h
b_h)-c_h\partial b_h]:-{1\over 12}+a^h,\label{gho}
\ee
which satisfies commutation relations analogous to
(\ref{commu})  with $a^m$ replaced by $a^h$ and 
a  central charge given by 
\cite{Friedan:1985ge,GSW} $c_h=1-3(2h-1)^2$. 
In (\ref{brst}), we also allowed for a normal ordering
constant $a$. The BRST charge 
depends on the normal ordering
constants only through  the combination $a^m+\sum a^h+a$.
Without loss of generality it is thus possible to choose 
$a^m=a^h=0$. 

The left moving BRST charge is similarly defined with its
corresponding left ghosts and antighosts except for $\gamma$ and
$\omega$ and the zero modes of $d$ and $e$
which are the same. This is due to the structure of $g$ 
as a sum of left movers and right movers and the fact that the
zero modes of $G$ and $\tilde G$ coincide.

The crucial property of $Q$ is its nilpotency.
The calculation of $Q^2$, using the full commutation relations
gives
\be
Q^2=- i{c_T \over 24}\int {d\sigma \over 2\pi}
\ (\partial^3c(\sigma))\ c(\sigma)-ia\int {d\sigma \over 2\pi}
c(\sigma)\partial c(\sigma),
\ee
where $c_T=c^m+\sum c_{h_i}$ is the total central charge
of the matter and ghost system which is given by $2D-26-2(D-N)$.
Thus the nilpotency of $Q$ requires  
that the total central charge $c_T$
and the normal ordering constant $a$ vanish. 
A similar conclusion is of course valid for the
left moving sector whose total central charge 
is given by $2D-26-2(N+2)$. The vanishing of
both central charges gives
$D=28$ and $N=13$.

\section{Spectrum}

The physical states are equivalence classes of states annihilated
by $Q$, two states being equivalent if their difference is
Q-exact. One has also to add some supplementary conditions 
originating from the ghosts zero modes. Notice that among the
ghosts $c, d$ and of course $\gamma$ together with their
antighosts have zero modes. This results in a degeneracy
of the ground state. By analogy with the usual
string theory \cite{GSW} we impose that 
the zero modes of the antighosts annihilate the physical states
\be
b_0|\Psi>=\tilde b_0|\Psi>=e_0|\Psi>=w|\Psi>=0.\label{zero}
\ee
This allows to impose correctly the zero mode constraints on the
physical states.

It will be convenient to use the following Fourier expansions
\ba
P_R(\sigma)&=&{p\over {2\pi}}+2{\mu}
\sum_{n\neq  0}
\beta_n e^{in\sigma},\ 
Q_R(\sigma)={q\over {2\pi}}+
{1\over {2\pi}}\sum_{n\neq 0}\alpha_n e^{in\sigma},\\
c(\sigma)&=&\sum_{n=-\infty}^{+\infty} c_n e^{i n\sigma},\ 
b(\sigma)=\sum_{n=-\infty}^{+\infty} b_n e^{i n\sigma}\\
T(\sigma)&=&\sum_{n=-\infty}^{+\infty} L_n e^{i n\sigma},\quad
G(\sigma)=
\sum_{n}G_ne^{in\sigma}\\
\hat P_R(\sigma)&=&-2i{\mu}\sum_{n\neq  0}
{\beta_n \over n}e^{in\sigma}.
 \ea
The Fourier coefficients of the fields satisfy
the commutation relations
\ba
\left[\beta^\mu_n,\alpha_m^\nu\right]=
m\eta^{\mu\nu}\delta_{n+m,0},\quad
\left\{c_n,b_m\right\}=\delta_{n+m,0},\quad
[q^\nu,\xi^\mu]=-i\eta^{\mu\nu},
\ea
the others  being zero.
We have
\ba
L^{(m)}_0&=& {p.q\over 4\pi \mu}
+\sum_{n\neq 0}:\beta_n.\alpha_{-n}:
+{7\over 3},\  
 L^{(m)}_m={p.\alpha_m\over{{4\pi\mu}}}
  +q.\beta_m+\sum_{n\neq 0,m}
 \beta_{n}.\alpha_{m-n},\nonumber\\
 L^{(c_h)}_0&=&\sum_n n :c_nb_{-n}: -{1 \over 12},\quad
 L^{(c_h)}_m=\sum_n\left[(h-1)m+n\right]c_nb_{m-n},\nonumber\\
  G_n&=&2\mu\left(
 \xi.{\beta_{n}}-{i}
 \sum_{m\neq 0}{\beta_m \over m}
 .{\beta_{n-m}}\right).\label{der} 
\ea
In equation (\ref{der}) we used $\beta_0=p/(4\pi \mu)$.

Notice that the total normal ordering constant coming from the
ghost sector is
$(D+4)/24=-2/3$ which when added to that of the matter sector
gives $1$.

If one defines similarily the left moving Fourier modes and
denotes them with a tilde then the total BRST charge, including
left and right contributions, reads
\ba
Q&=&\sum_{n}c_{n}(L_{-n}^{tot}-{L_{-n}^{(c)}\over 2})
+2{\mu}\left(\sum_{n\neq 0} c_{a'n}\beta^{a'}_{-n}
+\sum_{n\neq 0} c_{-n}\beta^{-}_{-n}\right)
+\sum_{n}d_{n}G_{-n}\nonumber\\
&+& 
\sum_{n}\tilde c_{n}(\tilde L_{-n}^{tot}-{\tilde L_{-n}^{(c)}
\over 2})
+2{\mu}\left(\sum_{n\neq 0} \tilde c_{an}\tilde \beta^{a}_{-n}
+\sum_{n\neq 0} \tilde c_{-n} \tilde \beta^{-}_{-n}\right)
+\sum_{n\neq 0}\tilde d_{n}\tilde G_{-n}\nonumber\\
&-&i{p^+\over {4\pi\mu}}[\sum_{n,m\neq 0}
c_n d_m b_{-(n+m)}^{-}+
\tilde c_n \tilde d_m \tilde b_{-(n+m)}^{-}]\nonumber\\
&+&
\gamma\left[g+{i\over 2\pi \mu}\left(\sum_{n\neq 0}(c_{-n}e_n+\tilde c_{-n}\tilde e_n)
+(c_0+\tilde c_0)e_0\right)\right]\nonumber\\
&+&d_0\left[G_0+i{p^+\over {4\pi\mu}}
\sum_{n\neq 0}c_nb_{-n}^{-}+\tilde c_n \tilde b_{-n}^{-}\right],
\ea
where
\be
g={{\xi^2}-1 }+\sum_{n\neq 0}
 \left({\beta_n.\beta_{-n}\over n^2}+
 {\tilde \beta_n.\tilde \beta_{-n}\over n^2}\right).
 \ee
Notice that as anticipated $\gamma$ and $d_0$ multiply
sums of right movers and left movers.
The vacuum state is defined by
\be
\alpha_n^\mu|0>=\beta_n^\mu|0>=c_n^{(h)}|0>=b_n^{(h)}|0>=0, 
\ \forall n>0
\ee
and for all the ghosts labelled by $h$ here.
The BRST operator acting on a vacuum state gives
\be
Q|0>=\left[c_0\left({p.q\over 4\pi\mu}+1\right)+d_0G_{ 0}+
\gamma\left(g+{i\over
2\pi\mu}c_0 e_0\right)\right]|0>.
\ee
If we now use the supplementary conditions
(\ref{zero}) then we get the following constraints on the
zero-mode part of the state
\be
(p.q+4\pi\mu)|0>=0,\quad \xi.p|0>=0,\quad (\xi^2-1)|0>=0.
\ee
These are precisely the conditions (\ref{conts})
defining a continuous spin
state \cite{wi}.

The additional zero mode constraints (\ref{zero}),
when anticommuted with the BRST charge $Q$ give the compatibility
conditions:
\ba
L_0^{tot}|\Psi>=0,\quad \tilde L_0^{tot}|\Psi>&=&0 \\
\left[G_0+i{p^+\over {4\pi\mu}}
\sum_{n\neq 0}c_nb_{-n}^{-}+\tilde c_n \tilde b_{-n}^{-}\right]
|\Psi>&=&0,\\
\left[g+{i\over 2\pi\mu}\sum_{n\neq 0}(c_{-n}e_n+\tilde c_{-n}\tilde e_n)
\right]|\Psi>&=&0.
\label{gmod}
\ea
Two states are equivalent if they differ by a Q exact
state of the form $Q|\Phi>$. The constraints (\ref{zero})
imply that $|\Phi>$ is not arbitrary 
but has to verify
\ba
b_0|\Phi>=0,\quad \tilde b_0|\Phi>=0,\quad
L_0^{tot}|\Phi>=0,\quad \tilde L_0^{tot}|\Phi>&=&0\\
w|\Phi>=0, \quad
[g+{i\over 2\pi\mu}\sum_{n\neq 0}(c_{-n}e_n+\tilde c_{-n}\tilde e_n)]|\Phi>&=&0\\
e_0|\Phi>=0, \quad [G_0+i{p^+\over{4\pi\mu}}
\sum_{n\neq 0}c_nb_{-n}^{-}+\tilde c_n \tilde b_{-n}^{-}]|\Phi>
&=&0.
\ea

Let us first determine the physical states with only matter excitations,
the ghosts being in the ground state.
The condition $Q|\psi>\otimes|0>=0$ gives for $|\psi>$, the state
in the matter sector,
\ba
\left(p.q+4\pi\mu(\sum_{n\neq 0}:\beta_n.\alpha_{-n}:+1)\right)|\psi>=0,
\quad L_n^{(m)}|\psi>=0,\ n>0\\
 \beta^{a'}_{n}|\psi>=0, \ n>0,
 \quad \beta^-_n|\psi>=0, \ n>0\ \,\\
 G_n|\psi>=0,\ n\geq 0,\quad  g|\psi>=0.
 \ea
 There are of course similar conditions for the left moving
 sector.
 The first condition determines
 the continuous spin parameter as $p.q=4\pi\mu(N-1)$, where 
 $N$ is the level of the state. Notice that for $N=1$,
 we have a reducible representation of the Poincar\'e group
 containing an infinite number of fixed helicity states.

Consider in more details the first level states
\be
|\psi>=(A_\mu\alpha_{-1}^\mu+B_\mu\beta^\mu_{-1})|0>,
\ee
The $L_0$ condition gives $p.q=0$, this state belongs to standard
helicity representations.
The $L_1$ condition yields
\be
{p.B\over 4\pi \mu}+q.A=0.
\ee
The $G_{ 0}$ and $G_{ 1}$ conditions give
\be
\xi.p=0,\quad (\xi-{i\over 4\pi \mu}p)^\mu A_\mu=0.
\ee
The $\beta_1$ conditions implies
\be
A_{+}=0=A_{a'}.
\ee
It remains to consider the $g$ condition
which gives
\be
({\xi^2 }-1)A_\mu=0,\quad
({\xi^2 }-1)B_\mu=A_\mu. \label{g1}
\ee
The nonzero solution of equation (\ref{g1})
is
\be
({\xi^2 }-1)=0,\quad A_\mu=0.
\ee
The first level physical state is thus
\be
|\psi>=B_\mu\beta^{\mu}_{-1}|0>,
\ee
with $p.B=0$. 
The  norm of the state is proportional to $A_\mu B^\mu$ 
which is zero. The standard helicity states are thus of zero norm.

Notice that the most constraining condition was $g|\psi>=0$
which implied alone the vanishing of $A_\mu$. 
We shall prove that this is also true for the higher level
states, that is the states annihilated simultaneously
by $L_0$ and $g$ do not contain $\alpha$ oscillators and are thus
of zero norm. 
Define $N_\alpha$ by $-\sum_{m=1}^{\infty}\alpha_{-m}.\beta_m/m$
and $N_\beta=-\sum_{m=1}^\infty\beta_{-m}.\alpha_m/m$.
$N_\alpha$ counts the number of $\alpha$ oscillators acting
on the ground state  and $N_\beta$ counts the number of $\beta$
oscillators, for example if 
\be
|\phi>= \alpha_{-n_1}^{\mu_1}\dots\alpha_{-n_n}^{\mu_n}|0>,
\ee
with all the $n_i$ strictly positive then
\be
N_{\alpha}|\phi>=n|\phi>.
\ee
Let $A$ be the part of $g$ depending on the right oscillators
$A=\sum_{m\neq 0}\beta_m.\beta_{-m}/m^2$. Then $A$ decreases the
number
of $\alpha$ oscillators by one and increases the number of 
$\beta$ oscillators by one:
\be
[N_\alpha,A]=-A,\quad [N_\beta,A]=A.
\ee
We want to show that an eigenstate of $A$
with a finite number of oscillators (an eigenstate of $L_0$)
has necessarily eigenvalue zero and does not contain $\alpha$
oscillators. For this decompose the eigenstate
as
\be
|\psi>=\sum_{n_\alpha=0}^N|\psi_{n_\alpha}> ,
\ee
with $|\psi_{n_\alpha}>$ a state containing $n_\alpha$ 
$\alpha$ oscillators acting on the vacuum, and 
$A|\psi>=a|\psi>$,
$a$ being the eigenvalue. Since $A$ decreases the number
of $\alpha$ oscillators by one, we have
\be
A|\psi>=\sum_{n_\alpha=1}^N|\chi_{n_\alpha-1}>,\label{aa}
\ee
where $|\chi_{n_\alpha-1}>=A|\psi_{n_\alpha}>$.
Equation (\ref{aa}) implies that $|\chi_{n_\alpha}>=
a|\psi_{n_\alpha}>$
for $n_\alpha=0,\dots N-1$ and $a|\psi_N>=0$.
Suppose that $a$ is not zero then $|\psi_N>=0$ and
$a|\psi_{N-1}>=|\chi_{N-1}>=A|\psi_N>=0$ and so $|\psi_{N-1}>$
vanishes and  all the other $|\psi_n>$ are zero.
So $a$ is necessarily $0$ and $|\psi>$ does not contain
any $\alpha$ oscillator.
We conclude  from the $g$ condition that
physical states are linear combinations of states of the form
\be
\beta_{-n_1}^{\mu_1}\dots \beta_{-n_N}^{\mu_N}|0>,
\ee
and these are all of zero norm.

Consider now the general state in the matter and ghosts Hilbert
space. The condition $g|\psi>=0$ is now replaced by
(\ref{gmod}) which implies again
that the state does not contain $\alpha$ oscillators 
and also that
the state does not contain $b$ nor $d$ oscillators.
The absence of $\alpha$ oscillators implies that 
 a physical state of non-zero norm is necessarily
of the form $|0>\otimes|\chi>$, where $|\chi>$ depends only
on ghosts excitations and $|0>$ is the ground state of the matter
sector. The physical state condition implies, in addition,
 that $|\chi>$ does
not contain $b$, $b^{a'}$, $b^{-}$ or  $e$ fermionic oscillators.
In turn, this implies that $|\chi>$ has a positive norm.
This completes the proof that there are no physical states with
a strictly negative norm.

\section{Conclusion}

Our main result is the absence of ghosts in the spectrum
and the presence of a physical state carrying 
the continuous spin
representation. 
These results do not seem to be dependent on the 
particular way we used to handle the bilocal constraint, that
is on its replacement by  a number of equivalent local ones 
in a given frame.
Notice in this respect that the physical spectrum we
obtained is Lorentz covariant. It would be interesting to treat
the bilocal constraint in a manifestly covariant way.
The results are however strongly dependent
on the requirement that physical states are annihilated by 
the zero modes of the antighosts which seems to be the key point 
in this BRST quantization. 
Our results prove the consistency of
the free theory. 
This is but the first step towards a consistent
theory with interactions where the role of the critical dimension
and the zero norm states should be very important.

\section*{Acknowledgements}
I am grateful to   X. Bekaert, K. Benakli, E. Dudas,
M. Petropoulos, 
B. Pioline, 
 S. Pokorski and G. Savvidy for
helpful discussions.

\end{document}